\documentstyle[12pt,bezier]{article}
\newcommand{\ssy}[5]{#1 19#4 {\it #2} {\bf #3} #5}
\title{Paradoxes of time travel}
\author{S. V. Krasnikov\thanks{Email: \it redish@pulkovo.spb.su}\\
The Central\\
Astronomical Observatory at Pulkovo, St. Petersburg,\\
196140, Russia}
\begin{document}
\maketitle
\begin{abstract}
Paradoxes that can supposedly occur if a time machine is created are
discussed. It is shown that the existence of trajectories of
``multiplicity zero'' (i.~e.~trajectories that describe a ball
hitting  its younger self so that the latter cannot fall
into the time machine) is not paradoxical by itself. This {\em
apparent paradox\/} can be resolved (at least sometimes) without any
harm to local physics or to the time machine. Also a simple model is
adduced for which the absence of {\em true} paradoxes caused by
self-interaction is proved.
\end{abstract}
\section{Introduction}
When one hears about time travel the first that comes to mind are
paradoxes.  In particular, a lot of paradoxes were described in science
fiction. They all are essentially versions of one fundamental
paradox, which can be formulated, for example, as follows.
\paragraph{The paradox.} Suppose a time traveller is going to kill
his younger self some time before the latter used the time machine.
His success would mean that he would never use the time machine and hence
would never commit the murder.  But as he would not be killed before
using the time machine nothing would prevent his travelling and
further killing his younger self, etc.
\paragraph{}To see that this paradox is, in fact, a
serious physical problem let us reformulate it. Consider an isolated
system which is prepared in some initial state $\eta(\tau_0)$ ($\tau$
is the proper time of the system). In a causal world a state of the
system at any $\tau=\tau_1>\tau_0$ is a function of $\eta(\tau_0)$.
This can be expressed (not rigorously) as:
     \begin{equation}
     \eta(\tau_1)=U_{\tau_0\to\tau_1}\big(\eta(\tau_0) \big)
     \label{USU}
        \end{equation}
Here $U$ is determined by local physical laws.  Assume, however, that
we build a time machine and our system gets (at some
$\tau^*:\;\tau_0<\tau^*<\tau_1$) into a region where closed timelike
curves exist. Now it can interact with its older self. The system at
$\tau_1$ may be ifluenced by itself at, say, $\tau_2:\;
\tau_2>\tau_1$. Instead of (\ref{USU}) we would have to write
     \begin{equation}
     \cases{
     \eta(\tau_1)=U_{\tau_0\to\tau_1}\big(\eta(\tau_0);
     \eta(\tau_2)\big)&\cr
     \eta(\tau_2)=U_{\tau_1\to\tau_2}\big(\eta(\tau_1)\big)&\cr}
     \label{PAR}
     \end{equation}
When a physical model we use is adequate, eq.~(\ref{USU}) has a
(unique) solution, but, in the general case, we do not know this for
system (\ref{PAR}). To {\em make\/} (\ref{PAR}) consistent is beyond
our power since $U$ is already fixed by conventional causal physics
(we consider only time machines like that from \cite{Mor,Pol}, where
causality violation is a manifestation of the global structure of
otherwise ``good'' spacetime; that is why we write the same $U$ in
(\ref{USU}) and (\ref{PAR})). And suppose that for some
$\eta(\tau_0)$ system (\ref{PAR}) has no solutions. How can this be
interpreted? What is the actual evolution of the system at
$\tau>\tau^*$? It is even hard to conceive where the answers  to these
questions may lie\footnote{Yet another reason to consider the paradoxes is
that they can be somehow connected with a hypothetical mechanism
``protecting causality'', which has long been looked for, but is not
yet at hand.}.
The above questions are somewhat academic until we find out whether
the paradoxes can really occur. The answer is not evident at all.
Apparent paradoxes like that with a traveller-suicide might be only
due to our overlooking  possible selfconsistent scenarios. For
example, the traveller might be injured instead of being killed. His
wound might be of such a nature that he would not be able any more to
shoot accurately enough and that is why later he would only injure his
younger self.
In Ref.~\cite{Ech} an attempt was made to find out whether paradoxes
present in a specific model.  A single perfectly elastic ball was
considered in a spacetime with a wormhole-based time machine. The
number of trajectories of the ball consistent with given initial data
(the {\em multiplicity\/} of the given trajectory\footnote{This term
is meaningful in the context since it turned out that the
multiplicity of a trajectory can be $>1$ and even infinite (see
Fig.~7 in \cite{Fri}). We shall see below that the nonexistence and
the nonuniqueness of trajectories may have the same source.} in terms
of \cite{Ech}) was evaluated.  The model, however, turned out to be too
complex and no definite answer was obtained to the question of
whether there exist trajectories of multiplicity zero.
Such a trajectory was found in
\cite{Ind}, where a simpler time machine (the Politzer spacetime, see
below) was considered. On the assumption that there are two perfectly
elastic balls with the different masses in this spacetime, it was
shown that their initial conditions in the past of the time machine
cannot be arbitrary.  This fact was interpreted in \cite{Ind} as an
argument against the existence of the time machine.
In the present paper we consider a model so simple that the following
becomes obvious:
\begin{enumerate}
\item  Trajectories with multiplicity zero (or with
multiplicity $>1$) are not at all unique to spacetimes with time
machines. Such trajectories appear whenever fixing the initial data
on the {\em partial\/} Cauchy surface one considers the region {\em
beyond\/} the Cauchy horizon.
\item The existence of a trajectory of multiplicity zero is not
paradoxical by itself. An {\em apparent\/} paradox like that found in
\cite{Ind} or that pursued in \cite{Ech} appears when one supplements
the abovementioned data with some {\em additional\/} conditions.
\end{enumerate}
Initial data inconsistent with {\em any\/} additional conditions
permissible in a given model can be called the {\em true\/} paradox.
We shall see that our model is free from true paradoxes, so the
question of whether there is anything paradoxical in time travel
remains unanswered.
\section{Apparent paradox}
In this section we construct  a situation similar to that from
\cite{Ind}. We somewhat simplify it to make it more open to
analysis.
Let us consider a world containing a time machine and populated only
by small elastic balls with the same masses. As a model of the time
machine (Fig.~1) we take Politser's spacetime \cite{Pol}. It is the
Minkowski plane where two cuts are made (say, along the segments
$t=\pm1, x\in(-1,1)$) and
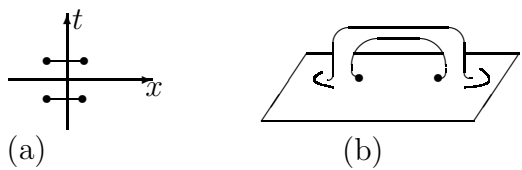
\begin{figure}[h,t,b]
\unitlength=0.250mm
\linethickness{0.4pt}
\begin{picture}(296.00,100.0)(-43.0,150.0)
\put(232.00,217.00){\oval(50.00,30.00)[t]}
\put(232.00,227.50){\oval(70.00,21.00)[t]}
\put(197.00,227.00){\line(0,-1){11}}
\put(267.00,227.00){\line(0,-1){10}}
\put(194.00,212.50){\oval(6.00,5.00)[rb]}
\bezier{116}(197.00,206.00)(180.00,209.00)(191.00,214.00)
\put(253.00,215.00){\oval(8.00,8.00)[rb]}
\put(271.50,217.00){\oval(9.00,12.00)[lb]}
\bezier{148}(267.00,206.00)(288.00,207.00)(275.00,216.00)
\put(211.00,211.00){\circle*{4.00}}
\put(253.00,211.00){\circle*{4.00}}
\put(257.00,216.00){\line(0,-1){2}}
\put(197.00,217.00){\line(0,-1){5}}
\put(183.00,224.00){\line(1,0){14}}
\put(267.00,224.00){\line(1,0){29}}
\put(296.00,224.00){\line(-2,-3){24}}
\put(208.00,224.0){\line(1,0){48}}
\put(183.00,224.00){\line(-2,-3){24}}
\put(272.0,188.00){\line(-1,0){113}}
\put(213.03,171.72){\makebox(0,0)[cc]{(b)}}
\put(55.78,183.68){\vector(0,1){61.47}}
\put(24.57,210.16){\vector(1,0){76.60}}
\put(61.45,242.31){\makebox(0,0)[cc]{$t$}}
\put(102.12,204.96){\makebox(0,0)[cc]{$x$}}
\put(44.43,220.09){\line(1,0){19.86}}
\put(64.29,199.76){\line(-1,0){19.86}}
\put(64.76,220.09){\circle*{4.0}}
\put(44.91,220.09){\circle*{4.0}}
\put(63.82,199.76){\circle*{4.0}}
\put(44.91,199.76){\circle*{4.0}}
\put(34.03,171.86){\makebox(0,0)[cc]{(a)}}
\put(211.50,217.50){\oval(9.00,11.00)[lb]}
\end{picture}
\caption{Politzer'z spacetime}
\end{figure}
\noindent after removing the points $t=\pm1,\,x=\pm1$ the upper bank of
each cut is glued to the lower bank of the other cut.  The resulting
manifold is a plane with handle and without two points (Fig.~1b).
World lines of balls we describe by inextendible polygonal lines
satisfying the following conditions
\begin{enumerate}
\item Each edge is a segment of a future-directed timelike straight
line.
\item Edges do not intersect, but they can meet in vertices.
\end{enumerate}
Now let us consider vertices, which are to describe collisions of
balls.
\begin{figure}[h,b,t]
\unitlength=1mm
\linethickness{0.4pt}
\begin{picture}(77.00,40.00)(24.0,65.0)
\put(50.00,74.00){\vector(1,1){24.00}}
 \put(55.00,73.00){\vector(1,2){13.00}}
\put(72.00,73.00){\vector(-1,1){23.33}}
\put(58.00,72.00){\circle*{0.30}}
\put(62.00,71.00){\circle*{0.30}}
\put(66.00,72.00){\circle*{0.30}}
\put(53.00,98.00){\circle*{0.30}}
\put(58.00,100.00){\circle*{0.30}}
\put(63.00,100.00){\circle*{0.30}}
\put(48.00,71.00){\makebox(0,0)[cc]{1}}
\put(54.00,70.00){\makebox(0,0)[cc]{2}}
\put(73.00,70.00){\makebox(0,0)[cc]{$n$}}
\put(46.00,99.00){\makebox(0,0)[cc]{1}}
\put(69.00,103.00){\makebox(0,0)[cc]{$(n-1)$}}
\put(77.00,101.00){\makebox(0,0)[cc]{$n$}}
\put(60.67,84.33){\circle*{1.0}}
\end{picture}
\caption{Same numbers correspond to same balls.}
\end{figure}
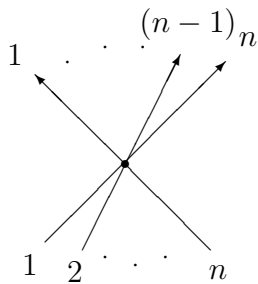
Since our spacetime is time-orientable we can distinguish
uniquely incoming and outgoing edges for any vertex. To make our
model complete we must formulate the rule determining the outgoing
edges from any set of the incoming ones. We adopt the following rule
(in what follows we number both incoming and outgoing edges from left
to right):
\begin{itemize}
\item[3.]
\begin{itemize}
\item[(a)] Each vertex with $n$ incoming edges has $n$ outgoing edges.
\item[(b)] The $k\,$th  outgoing edge has the same direction
as the $(n-k+1)\,$th incoming edge. So the vertex looks like an
intersection of $n$ straight lines.
\item[(c)] $k\,$th incoming and $k\,$th outgoing edges are deemed
consequent parts of the same world line.
\end{itemize}
\end{itemize}
For $n=2$ this rule is equivalent to the statement that our balls are
perfectly elastic and distinguishable. For larger $n$ we must adopt
the rule 3 if we want the coordinates of the balls after collision to
depend continuously on those before the collision.
The rules 1--3 allow one to trace the history of balls if one fixes
each ball by assigning at some $t$ its coordinate and velocity or
a part of its trajectory. Now we can construct a ``trajectory of
multiplicity zero''.
Consider the two bold segments $(A_1,B_1)$ and $(A_2,B_2)$ in
Fig.~3a.  Let them be parts of the world lines of two balls.
%
%
 %
\begin{figure}[h,t,b]
\unitlength=0.6mm
\linethickness{0.4pt}
\begin{picture}(152.00,65.00)(-10,40)
{\linethickness{0.2pt}
\put(15.00,100.00){\line(1,0){35.00}}
\put(15.00,65.00){\line(1,0){35.00}}}
\put(28.00,100.00){\line(3,-4){33.0}}
\put(6.00,96.00){\line(3,-4){23.25}}
\put(37.00,100.00){\line(-3,-4){33.67}}
\put(61.10,96.80){\line(-3,-4){23.8}}
\put(15.00,100.00){\circle*{1.60}}
\put(15.00,65.00){\circle*{1.60}}
\put(50.00,100.00){\circle*{1.60}}
\put(50.00,65.00){\circle*{1.60}}
\put(116.00,100.00){\line(1,0){35.00}}
{\linethickness{0.2pt}
\put(116.00,65.00){\line(1,0){35.00}}
\put(15.00,65.00){\line(1,0){35.00}}}
\put(129.00,100.00){\line(3,-4){33.0}}
\put(107.00,96.00){\line(3,-4){23.25}}
\put(138.00,100.00){\line(-3,-4){33.67}}
\put(162.10,96.80){\line(-3,-4){23.8}}
\put(116.00,100.00){\circle*{1.60}}
\put(116.00,65.00){\circle*{1.60}}
\put(151.00,100.00){\circle*{1.60}}
\put(151.00,65.00){\circle*{1.60}}
\put(122.00,65.00){\line(0,1){35.00}}
\put(142.00,65.00){\line(0,1){35.00}}
\put(30.00,45.00){\makebox(0,0)[cc]{(a)}}
\put(131.00,45.00){\makebox(0,0)[cc]{(b)}}
\put(21.00,82.00){\makebox(0,0)[cc]{$F$}}
\put(46.00,82.00){\makebox(0,0)[cc]{$D$}}
\put(37.00,94.00){\makebox(0,0)[cc]{$G$}}
\put(28.00,104.00){\makebox(0,0)[cc]{$E$}}
\put(28.00,61.00){\makebox(0,0)[cc]{$E$}}
\put(37.00,61.00){\makebox(0,0)[cc]{$C$}}
\put(37.00,104.00){\makebox(0,0)[cc]{$C$}}
\put(64.00,101.00){\makebox(0,0)[cc]{$D'$}}
\put(6.00,101.00){\makebox(0,0)[cc]{$F'$}}
\put(7.00,52.00){\makebox(0,0)[cc]{$A_1$}}
\put(59.00,52.00){\makebox(0,0)[cc]{$B_1$}}
\put(11.00,60.00){\makebox(0,0)[cc]{$A_2$}}
\put(64.50,60.00){\makebox(0,0)[cc]{$B_2$}}
\thicklines
\put(3.00,55.00){\line(3,4){6.00}}
\put(56.00,63.00){\line(3,-4){6.00}}
\put(3.30,55.00){\line(3,4){6.00}}
\put(56.30,63.00){\line(3,-4){6.00}}
\end{picture}
\caption{The apparent paradox and its resolutions.}
\end{figure}
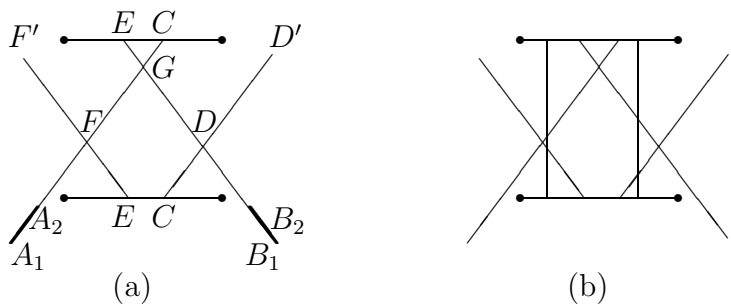
\paragraph{Statement.} There is no evolution of two balls consistent
with these initial conditions.
\paragraph{Proof.} Suppose such evolution exists. It
follows then from the rule 3 that the world line of the left ball
cannot lie to the right of the ray $FF'$. Similarly, the world line
of the right ball cannot lie to the left of the ray $DD '$.  So
the balls cannot collide. Also neither of them can hit
its younger self as they are bounded away from the segment
$\{y=1,\,-1<x<1\}$ by $FF'$ and $DD '$. But this implies that the
balls will reach unobstructed the point $G$.  Here they must meet ---
a contradiction.
Note that what we have obtained is actually a version of the paradox
with a traveller killing his younger self.
\section{Resolution}
The resolution of the paradox lies in the term ``initial conditions''.
Let us introduce two new terms. We shall call conditions set on any
partial Cauchy surface (which is a  connected achronal hypersurface
without boundary) {\em partial initial conditions}, PIC. And anything
that, taken together with the local laws, determines uniquely
evolution of balls in a spacetime under consideration we shall call
{\em complete initial conditions}, CIC.
To begin with let us consider the Minkowski plane (Fig.~4) and the
partial Cauchy surface $S:\;t^2-x^2=1$ in it.
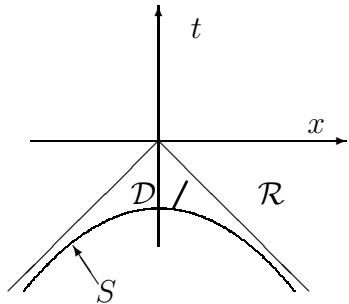
\begin{figure}[h,b]
\unitlength=1mm
\linethickness{0.4pt}
\begin{picture}(75.00,40.00)(0,78)
\bezier{228}(32.00,80.00)(50.00,102.00)(68.00,80.00)
\put(50.0,100.0){\line(-1,-1){20}}
\put(50.0,100.00){\line(1,-1){20}}
\put(50.00,86.00){\vector(0,1){32.00}}
\put(33.00,100.00){\vector(1,0){42.00}}
\put(55.00,115.00){\makebox(0,0)[cc]{$t$}}
\put(71.00,102.00){\makebox(0,0)[cc]{$x$}}
\put(42.00,81.00){\vector(-2,3){3.33}}
\thicklines
\put(52.00,91.0){\line(1,2){1.8}}
\put(43.00,80.00){\makebox(0,0)[cc]{$S$}}
\put(65.00,93.00){\makebox(0,0)[cc]{$\cal R$}}
\put(48.00,93.00){\makebox(0,0)[cc]{$\cal D$}}
\end{picture}
\caption{The bold segment is a part of a single world line
intersecting $S$. It can be used as PIC.}
\end{figure}
We can assign arbitrary
conditions to balls on this surface and they will completely
determine evolution of these balls inside the domain $\cal D$ lying in
the past cone of the origin of coordinates. ($\cal D$, which by
definition is a region where all past inextendible causal curves
intersect $S$, is called the {\em Cauchy development\/} of $S$.) However,
to fix evolution of the balls in the region $\cal R$, i.~e.~outside
$\cal D$, we need some additional conditions.  In other words, the
PIC set on $S$ are CIC in its Cauchy development, but cease to be
CIC as soon as we intersect the Cauchy horizon (which is the boundary
of $\cal D$).
An important fact is that additional conditions we need for
completing CIC affect the multiplicity of the trajectory determined
by the PIC. Consider, for example, the situation depicted in Fig.~4
and assume we consider massless balls as well. As an additional
condition one can require that there should be $m$ balls in $\cal R$.
Then for $m=0$ one obtains a trajectory of multiplicity zero, for
$m=1$, a unique trajectory, and for $m=2$, a trajectory of the
infinite multiplicity.
The paradox cited in the previous section is just of the same nature.
Conditions set on the partial Cauchy surface $t=t_0<-1$ cease to be
CIC as soon as we consider the region beyond the Cauchy horizon. The
CIC here consist of the PIC and the requirement that there are {\em
only two\/} balls in the spacetime.  This requirement follows neither
from PIC, nor from the local laws 1--3 (nor from common sense or
intuition). It is an independent constraint and no wonder that it
comes into conflict with PIC. If we abandon this constraint, we
get infinitely many solutions satisfying both the equations of
motions (1--3) and the PIC at $t<0$.  One of these solutions is shown
in Fig.~3a, where the closed polygonal lines $(EFGE)$ and $(CDGC)$
are interpreted now as the world lines of two additional balls.
We can see now the weak point in suggestion (2). A system isolated to the 
past of the time
machine can cease to be isolated
beyond the horizon and we cannot completely control its interaction
there.
\paragraph{Note.} The parallels between the causal and acausal cases
discussed in this section can be made all the more evident if we
``roll up'' the part $x>t$ of the Minkowski plane into the Misner
space (see \cite{Haw} for details). The past cone of the origin of
coordinates in this process will go into the causal part of the
cylinder and the part of $\cal R$ will just form a time machine.
\section{Discussion}
The above solution of the paradox is typical for Politzer's spacetime
containing only pointlike perfectly elastic balls with the same
masses.
\smallskip{\em
There are no true paradoxes in such a world. For any initial conditions in
the past of the time machine there exist infinitely many solutions
satisfying both these conditions and the equations of motion.}
\smallskip To find these solutions, take a set of straight lines
satisfying the PIC.  Mutual intersections break these lines into
portions (segments and rays). Number these portions from left to
right for any $t$.  Interpreting each portion as a part of the world
line of the ball with the same number we get a desired solution.
Arbitrarily many other solutions can be obtained simply by drawing
additional vertical lines (see Fig.~3b).
Unfortunately, this method of finding a solution fits only the
considered model, which, perhaps, is free from paradoxes only owing
to its simplicity. To make the next step one should, possibly,
consider various models available from this one  by
introducing different sorts of balls and arbitrary preassigned
interaction instead of (3) (it is unlikely that our choice of the
spacetime leads to noticeable loss of generality).  It seems that the
only reasonable restriction on the choice of interaction is the
requirement that it must respect the symmetries of the spacetime.
That is, if one admits the vertex Fig.~5a, one must admit also the
vertices 5b,c.
\begin{figure}
\unitlength=1mm
\begin{picture}(117.00,35.00)(0.0,76.0)
\put(20.0,100.0){\line(-2,-3){10}}
\put(20.00,101.00){\circle*{0.30}}
\put(20.00,104.00){\circle*{0.30}}
\put(20.00,107.00){\circle*{0.30}}
\put(20.00,110.00){\circle*{0.30}}
\put(60.0,96.00){\line(2,3){10}}
\put(60.00,94.00){\circle*{0.30}}
\put(60.00,91.00){\circle*{0.30}}
\put(60.00,88.00){\circle*{0.30}}
\put(60.00,85.00){\circle*{0.30}}
\put(7.00,90.00){\vector(3,4){5.33}}
\put(32.00,90.00){\vector(-3,4){5.33}}
\put(25.00,104.00){\vector(0,1){9.00}}
\put(94.0,100.0){\line(-1,-5){3}}
\put(95.00,101.00){\circle*{0.30}}
\put(97.00,103.00){\circle*{0.30}}
\put(99.00,105.00){\circle*{0.30}}
\put(101.00,107.00){\circle*{0.30}}
\put(86.00,87.00){\vector(1,4){2.67}}
\put(112.00,87.00){\vector(-1,1){8.00}}
\put(103.00,103.00){\vector(1,1){7.00}}
\put(20.00,75.00){\makebox(0,0)[cc]{(a)}}
\put(60.00,75.00){\makebox(0,0)[cc]{(b)}}
\put(94.00,75.00){\makebox(0,0)[cc]{(c)}}
\put(6.00,94.00){\makebox(0,0)[cc]{$ V_1$}}
\put(33.00,94.00){\makebox(0,0)[cc]{$ V_2$}}
\put(83.00,94.00){\makebox(0,0)[cc]{$ V'_1$}}
\put(111.00,94.00){\makebox(0,0)[cc]{$ V'_2$}}
\put(29.00,109.00){\makebox(0,0)[cc]{$ V_3$}}
\put(117.00,109.00){\makebox(0,0)[cc]{$ V'_3$}}
\thicklines
\put(20.0,100.0){\line(2,-3){10}}
\put(60.0,96.0){\line(-2,3){10}}
\put(94.0,100.00){\line(1,-1){15}}
\end{picture}
\caption{$V'_i$ are obtained from $V_i$ by some boost.}
\end{figure}
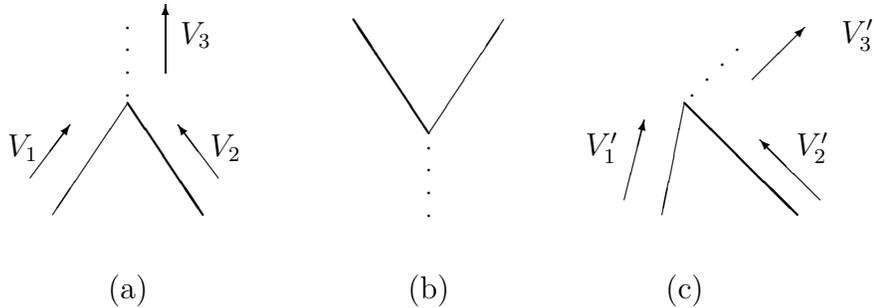
It should be particularly emphasized that the
questions posed in the Introduction will face us as soon as we find a
true paradox for {\em any\/} such model no matter how detailed and
realistic it is. For, suppose that a paradox presents in some model,
but disappears if one complements the model with, say, ``back
reaction'' (i.~e.~adds Einstein equation to it). This would mean
that there is some intimate connection between Einstein equations and
the problem of existence of evolution for some quite abstract
systems. Such a discovery would be every bit as surprising as the
paradox itself.
On the other hand the absence of true paradoxes in these generalized
models would strongly suggest that they do not occur at all.
If this is the case and any paradox can be (and must be) resolved in
the manner shown above, one might obtain quite a bizarre picture. In
science fiction terms it looks like the following.  A potential
suicide gets a wormhole. He checks (for example, by travelling
through it) that the throat of the wormhole is empty. By moving the
mouths in the appropriate way \cite{Mor} he transforms the wormhole
into a time machine. Then he loads his gun and enters the mouth. And
here, all of a sudden, he meets a policeman, who disarms him
preventing the murder ({\em and\/} the paradox).
We know that neither the existence of the policeman, nor his
interaction with the traveller contradict any local physical laws.
So, if we can find an appropriate world line for the policeman (it
must not, in particular, intersect the Cauchy horizon), we will have
to acknowledge that, however {\em strange\/}, this situation is not
{\em paradoxical.}

\end{document}